\documentclass{cpbtex}
\usepackage{graphicx}
\usepackage{CJK} 
\usepackage{tikz}
\usetikzlibrary{shapes.geometric, arrows.meta, positioning, decorations.pathmorphing}
\usepackage{float}
\usepackage{pgfplots}
\pgfplotsset{compat=1.9}
\usepackage{xurl}
\usepackage{xcolor}

\begin{document}
	\begin{CJK*}{UTF8}{gbsn} 
		\title{Hierarchical Progressive Optimization for Multi-Qubit Pauli Noise Modeling}
		
		\author{
			Xiangyu Ge(葛翔宇)$^{1}$, 
			Jiafei Ge(葛佳菲)$^{2}$, 
			Shengmei Zhao(赵生妹)$^{1}$, 
			Le Wang(王乐)$^{1}$,\\
			and Anqi Zhang(张安琪)$^{1,}$\thanks{{Corresponding author. E-mail: zhangaq@njupt.edu.cn}}\\[6pt]
			$^{1}$Nanjing University of Posts and Telecommunications, Nanjing 210003, China\\
			$^{2}$School of Basic Medical Sciences \& School of Public Health,\\
			Faculty of Medicine, Yangzhou University, Yangzhou 225009, PR China
		}
		
		\date{\today}
		
		\maketitle
			
		\begin{abstract}

		Quantum Noise Characterization (QNC) is indispensable for benchmarking and mitigating errors in Noisy Intermediate-Scale Quantum (NISQ) devices. However, traditional Quantum Process Tomography (QPT) suffers from an exponential parameter explosion, severely hindering its scalability. 
		In this paper, we propose a Hierarchical Progressive Optimization (HPO) framework to efficiently extract high-order spatial crosstalk in multi-qubit systems. The complexity analysis shows that the combinatorial projection mask reduces the \textcolor{black}{required number of Pauli transfer matrix (PTM) elements} from $\mathcal{O}(16^N)$ to $\mathcal{O}(N^2 3^N)$. 
		{Numerical simulations on a 10-qubit HHL circuit achieve a fidelity of $0.9381$ with the HPO method, compared to $0.7431$ obtained using global depolarizing-noise mitigation.}
		\end{abstract}

		\maketitle

		\begin{center}
			\begin{minipage}{0.95\linewidth}
			\\[2pt]
			\textcolor{black}{\textbf{PACS:} 03.65.Wj, 03.67.Lx, 03.65.Yz, 03.67.-a}
			\end{minipage}
		\end{center}
		\section{Introduction}
		
		In the Noisy Intermediate-Scale Quantum (NISQ) era, hardware imperfections and decoherence cause the execution outcomes of quantum algorithms on physical devices to deviate significantly from ideal simulations \cite{Xue2021}. Such quantum noise not only degrades the fidelity of multi-qubit states in operations like quantum teleportation \cite{Wang2024}, but also poses a critical bottleneck for quantum metrology \cite{Lei2024} and error mitigation tasks \cite{Li2025, Li2017}. 
		Recent studies on the fundamental limits of quantum error mitigation further emphasize that without highly accurate physical noise models, achieving practical quantum advantage remains extremely difficult \cite{Quek2024}. Therefore, an accurate and scalable quantum noise model serves as the indispensable bridge connecting theoretical quantum algorithms with real-world physical hardware, fundamentally impacting the evaluation of quantum architectures and performance prediction.
		
		To characterize these complex quantum channels, various tomographic methods have been proposed. Full Quantum Process Tomography (QPT) and Pauli Transfer Matrix (PTM) reconstruction provide a mathematically exhaustive description of the noise maps \cite{Stanchev2024, Roncallo2023}. Although advanced classical shadow techniques have been introduced to alleviate measurement overhead \cite{Levy2024}, these full-characterization methods still severely lack scalability due to exponential parameter explosion, requiring $\mathcal{O}(16^N)$ measurements for an $N$-qubit system. {At the other extreme}, simplified phenomenological models (e.g., independent depolarizing channels) are highly scalable but inherently fail to capture the ubiquitous spatial crosstalk in contemporary quantum processors. Early studies identified crosstalk as a primary source of multi-qubit errors \cite{Sarovar2020}, significantly manifesting in both readout processes \cite{Maciejewski2020, Maciejewski2021} and microwave-driven gate operations \cite{Wang2022}. Recent physical experiments explicitly demonstrate that simultaneous gate operations on superconducting qubits induce profound irreducible multi-body crosstalk ($\mathcal{E}_{ij} \neq \mathcal{E}_i \otimes \mathcal{E}_j$) \cite{Zhao2025PRL}, which can only be diagnosed via complex out-of-time-order correlators (OTOCs) \cite{Zhang2024} and suppressed by specialized hardware shielding structures \cite{Tang2024}. Consequently, existing methods either suffer from parameter explosion or are overly simplified to express true physical spatial coupling. In particular, there is a distinct lack of a recursive extension mechanism that clearly answers ``which parameters can be directly inherited from low-qubit systems, and {which high-order parameters require additional optimization}.''
		
		The core contradiction in current noise modeling lies in the exponentially growing dimensionality of full PTMs, in contrast to the severely constrained measurement budget on physical quantum hardware. Therefore, a structured modeling approach that can utilize low-qubit samples and recursively extend to high-qubit systems is critically required. To break through this bottleneck, recent studies have actively explored approximation schemes based on physical spatial structures. For instance, graph attention networks have been utilized to capture nearest-neighbor interactions in many-body systems \cite{Lin2024}, and local receptive fields have been applied to truncate redundant long-range correlations in quantum error correction codes \cite{Ji2025, Li2025surface}. These works reveal that discarding the global fully-connected assumption and leveraging locality in quantum systems has emerged as a critical strategy to mitigate the dimensionality curse. However, at the hardware level, mechanisms such as maintaining low-noise stability \cite{Huang2025} and leveraging opposite-sign anharmonicity \cite{Zhao2022} or crosstalk-robust quantum control \cite{Zhou2023} reveal that quantum noise exhibits stable locality accompanied by fragile high-order correlations. Few studies have explicitly analyzed information retention capacity under Pauli weight and Hamming distance constraints. Specifically: (1) the dimension of the Pauli basis grows exponentially with the number of qubits; (2) quantum noise comprises not only single-body errors but also two-qubit crosstalk and multi-body correlations; and (3) hardware measurement costs are prohibitively high and sample sizes are strictly limited, precluding reliance on full parameter reconstruction. As the system scales, high-weight multi-qubit Pauli correlations are highly susceptible to decoherence and physically collapse toward zero---a phenomenon we define as \textit{Pauli dilution}. Ultimately, quantum noise intrinsically presents a duality of robust ``local coupling features'' and fragile ``high-order correlation effects.'' This physical constraint requires that a scalable, reliable noise model avoid both overly simplistic uncorrelated approximations and unrestricted full-dimensional expansion in the exponential Hilbert space.
		
		To address these challenges, we propose a Hierarchical Progressive Optimization (HPO) framework for sparse Pauli noise modeling, tailored for characterizing multi-qubit quantum circuits. Unlike traditional approximation methods that optimize a truncated global space simultaneously, our framework physically decouples the noise into a local baseline and high-order residuals. Specifically, we first establish a 2-qubit foundational model constrained by a Hamming distance mask ($d \le 2$) to capture dominant local errors. When extending to $N$-qubit systems ($N > 2$), the framework strictly freezes the pre-trained low-order parameters inherited from the baseline.
		
		The contributions of this paper are {fourfold}: 
		1) a Hamming distance mask is introduced for sparse Pauli noise modeling, which preserves dominant local errors and filters irrelevant Pauli correlations.
		2) a hierarchical freezing mechanism is developed, in which the pre-trained low-order parameters are frozen when the system is extended to larger multi-qubit systems.
		3) a topology-aware parameter sharing scheme is constructed, in which the optimized 2-qubit foundational model is reused on the global topology graph.
		4) a recursive progressive optimization framework is established, in which the noise model is learned from the foundational baseline to the high-order residuals.
		
		The proposed HPO {framework} provides a reliable, scalable paradigm for evaluating high-fidelity multi-qubit gate schemes \cite{Zhao2024}, robust quantum gate optimizations \cite{Wang2025}, and reliable conversion schemes for hybrid entangled states in error-predicted devices \cite{Du2025}.





		\section{Proposed Sparse Pauli Noise Modeling Methodology}
		
\begin{figure*}[htbp]
			\centering
			\includegraphics[width=\columnwidth]{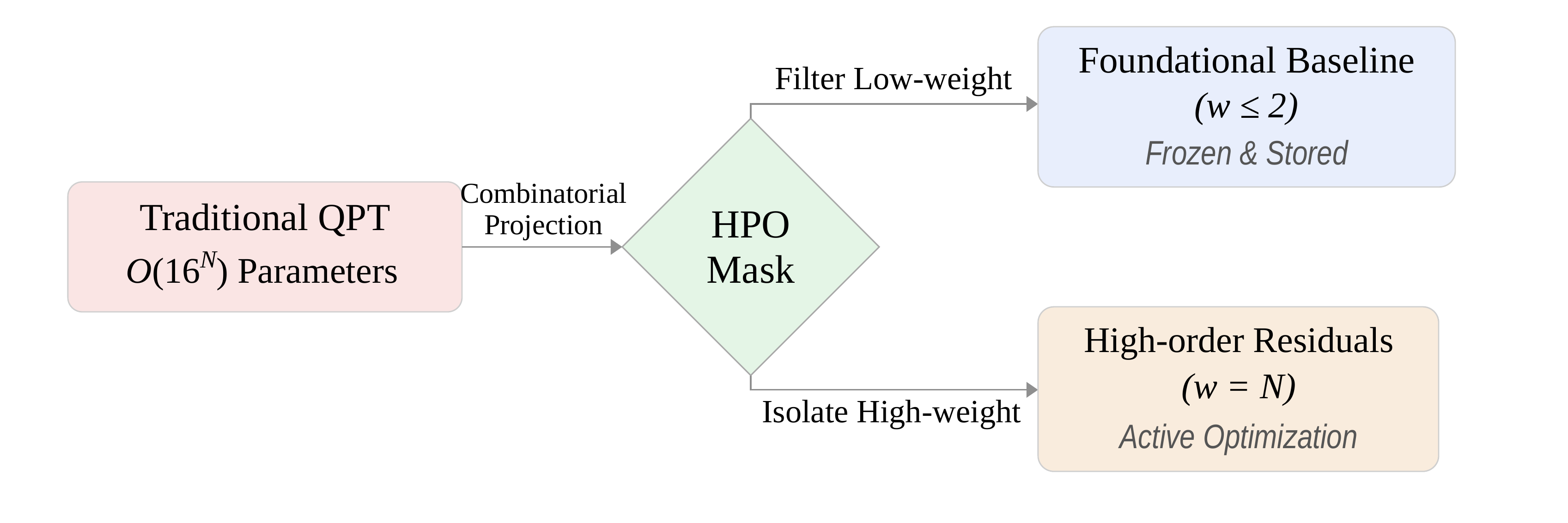}
			\caption{Schematic diagram of the proposed Hierarchical Progressive Optimization (HPO) framework. By introducing a deterministic combinatorial mask, the framework successfully decouples the intractable multi-body interactions into a frozen foundational baseline and an active optimization subspace, effectively avoiding the parameter explosion.}
			\label{fig_framework}
		\end{figure*}

		To address the exponential parameter explosion inherent in full Quantum Process Tomography (QPT) while accurately capturing spatial crosstalk, we propose a Hierarchical Progressive Optimization (HPO) framework. The core philosophy is to transition from global unconstrained optimization to a strict physics-informed subspace search. As schematically illustrated in Fig.~\ref{fig_framework}, the detailed formulation is presented in the following four subsections.
		
		\subsection{Mathematical Formulation of the 2-qubit Foundational Baseline}
		The HPO framework stems from systematic characterization of the elementary two-qubit coupling unit, the core building block for large-scale quantum systems. For an $N$-qubit system, the complete operator space is spanned by the Pauli basis $\mathcal{P}^{\otimes N} = \{I, X, Y, Z\}^{\otimes N}$. Under the Markovian approximation, any completely positive and trace-preserving (CPTP) quantum channel $\mathcal{E}$ can be uniquely mapped into a real-valued Pauli Transfer Matrix (PTM) $\mathcal{R}$, with its elements defined by the Hilbert-Schmidt inner product {\cite{Roncallo2023}}:
		\begin{equation}
			\mathcal{R}_{ij} = \frac{1}{2^N} \text{Tr}\left[ P_i \mathcal{E}(P_j) \right].
		\end{equation}
		For a localized 2-qubit subsystem, the full PTM resides in a $16 \times 16$ dimensional continuous space, containing {256 matrix entries}. However, in superconducting architectures, physical noise is predominantly highly localized, primarily driven by single-qubit relaxation ($T_1$), dephasing ($T_2$), and adjacent static ZZ-crosstalk. Simultaneously optimizing all {256 PTM entries} with limited experimental measurements may cause severe overfitting to unphysical high-order entangled noise.
		
		To enforce physical sparsity and isolate these dominant local errors, we introduce a Hamming distance mask $M^{(2)}$. Let $w(P)$ denote the Pauli weight (the number of non-identity operators in the string), and $D_H(P_i, P_j) = \sum_{k} \mathbb{I}(P_i^{(k)} \neq P_j^{(k)})$ denote the Hamming distance underlying the transition probability. This mask plays a central role in our framework: superconducting architectures demonstrate that simultaneous physical errors beyond two-body Markovian interactions are exponentially suppressed \cite{Sheldon2016PRA}. By restricting the Hamming distance, the mask effectively filters out unphysical noise artifacts and confines parameter dynamics to local physical processes. The foundational model is structurally constrained as:
		\begin{equation}
			\mathcal{R}_{\text{base}}^{(2)} = \mathbb{I} + \left( \Delta \mathcal{R} \odot M^{(2)} \right),
		\end{equation}
		\begin{equation}
			\text{s.t.} \quad M^{(2)}_{ij} = 
			\begin{cases} 
				1, & \text{if } \big(w(P_i) \le 2 \lor w(P_j) \le 2\big) \land D_H(P_i, P_j) \le 2 \\
				0, & \text{otherwise}
			\end{cases}
		\end{equation}
		where $\mathbb{I}$ represents the ideal identity channel, and $\odot$ denotes the Hadamard product. By applying this hard-truncation mask, this foundational phase precisely captures essential single-qubit decoherence and neighboring two-qubit crosstalk. The optimized parameters obtained in this stage form a stable, physically grounded baseline for subsequent high-order extensions.
		
		\subsection{Topology-Aware Recursive Extension via Kronecker Mapping} 
		While the foundational model accurately captures local dynamics, unconstrained full-dimensional PTM reconstruction ($\mathcal{O}(16^N)$) for an $N$-qubit processor ($N > 2$) incurs an exponential parameter explosion, making it experimentally infeasible.
		To address this, we project the optimized foundational 2-qubit parameters onto the macroscopic system topology graph $G=(V, E)$, where $V$ denotes the physical qubits and $E$ denotes valid coupling edges.
		
		We define a physical operator $\Lambda_{e}(\cdot)$ that maps a localized 2-qubit PTM on a specific edge $e=(u,v)$ into the global $N$-qubit Hilbert space. 
		Under the assumption that spontaneous long-range crosstalk is absent without physical graph connectivity, the global sparse base PTM $\mathcal{\tilde{R}}^{(N)}_{\text{base}}$ is recursively constructed from a tensor product sequence:
		\begin{equation}
			\mathcal{\tilde{R}}^{(N)}_{\text{base}} = \bigotimes_{e=(u,v) \in E} \Lambda_{e} \left( \mathcal{R}^{(2)}_{\text{base}, e} \right), \quad \text{where } \Lambda_{e}(\mathcal{R}) = \mathcal{R} \otimes I_{V \setminus \{u,v\}}.
		\end{equation}
		This recursive spatial extension projects the empirically verified local error characteristics onto the expansive $N$-qubit space. By strictly padding the non-interacting spectator qubits with the identity operator $I$, we ensure that no artificial, non-local crosstalk artifacts are mathematically generated during the scaling process. This step acts as a mean-field approximation of the global noise landscape.

		
		\subsection{Hierarchical Progressive Optimization with Gradient Projection}
		Although the recursively extended base model efficiently captures topological errors, it inherently underfits genuine high-order multi-body correlations that emerge during simultaneous multi-qubit gate operations. 
		Our approach centers on the Hierarchical Progressive Optimization (HPO) framework, which characterizes weak spatial crosstalk via rigorous gradient projection.
		
		We implement a topology-aware parameter sharing strategy to decouple the loss landscape. For any Pauli basis where the weight $w(P) < N$, the structural parameters are strictly frozen to their pre-trained values inherited from lower-order stages, denoted as $\mathcal{W}_{\text{frozen}}$. To capture the fragile $N$-body correlations without suffering from the barren plateau problem, {the theoretical PTM residual space is confined to the $N$-th-order mask-selected subspace}.
		
		We mathematically define the dynamic progressive mask $M^{(N)}$ as a hard physical truncation filter:
		\begin{equation}
			M^{(N)}_{ij} = 
			\begin{cases} 
				1, & \text{if } \big(w(P_i) = N \lor w(P_j) = N\big) \land D_H(P_i, P_j) \le 2 \\
				0, & \text{otherwise}
			\end{cases}
		\end{equation}
		The effective global transformation matrix for the $N$-th order characterization is formulated as $\mathcal{R}^{(N)}_{\text{eff}} = \mathbb{I} + \mathcal{W}_{\text{frozen}} + \big( \Delta \mathcal{R}_{\text{res}} \odot M^{(N)} \big)$. Crucially, during the backpropagation process, the gradient of the tomographic loss function $\mathcal{L}$ with respect to the incremental residual tensor $\Delta \mathcal{R}_{\text{res}}$ is mathematically projected onto the defined subspace:
		\begin{equation}
			\theta_{t+1} = \theta_t - \eta \left( \nabla_{\Delta \mathcal{R}} \mathcal{L} \odot M^{(N)} \right).
		\end{equation}
		{The element-wise projection restricts the theoretical PTM residual entries to the $N$-th-order mask-selected space.}
		This explicit orthogonal projection prevents the optimizer from wandering into the redundant $16^N$ flat manifold, thereby successfully avoiding complex, non-convex local minima and ensuring stable, machine-precision convergence for high-order residuals.
		
		\subsection{Combinatorial Complexity Analysis and Dimensionality Reduction}
		The primary bottleneck of full Quantum Process Tomography (QPT) is the curse of dimensionality. Let $\mathcal{P}^{\otimes N}$ be the $N$-qubit Pauli group. \textcolor{black}{The required PTM-element count scales exponentially as} $\mathcal{D}_{\mathrm{full}} = 16^N$. The HPO framework systematically mitigates this issue through two-phase dimensionality reduction.
		
		\textbf{Phase 1: Linear Scaling of the Foundational Baseline.} 
		The localized foundational model restricts active parameters strictly to physical coupling edges $e \in E$. For a specific edge, the distance mask $d \le 2$ bounds the parameter count to a maximum of $16^2 = 256$. By mapping this onto the global topology $G=(V, E)$, the base network complexity is bounded by the linear sum over all physical edges:
		\begin{equation}
			\mathcal{D}_{\text{base}} \le \sum_{e \in E} 256 = 256|E|.
		\end{equation}
		For contemporary superconducting processors with bounded-degree connection topologies (e.g., a 1D chain where $|E|=N-1$), the number of edges scales linearly with the number of qubits. Thus, the optimization complexity of the foundational baseline exhibits \textbf{linear growth {for bounded-degree connection topologies}}: $\mathcal{D}_{\text{base}} \sim \mathcal{O}(N)$.
		
		\textbf{Phase 2: Combinatorial Bound of Progressive Residuals.} 
		During the $N$-th stage of progressive optimization, attempting to capture full multi-body correlations would reintroduce the $\mathcal{O}(16^N)$ scaling. However, the progressive mask $M^{(N)}$ isolates only the critical $N$-th order subspace, defined by the logical constraint: $\big(w(P_i) = N \lor w(P_j) = N\big) \land D_H(P_i, P_j) \le 2$.
		
		We mathematically derive the exact active parameter count, $K_{\text{res}}^{(N)}$, using combinatorial analysis. Let $\mathcal{S}_N$ denote the set of Pauli strings with a full weight of $w=N$. Since these strings contain no identity operators ($I$), they are composed entirely of $X, Y, Z$, yielding $|\mathcal{S}_N| = 3^N$.
		
		\textcolor{black}{For each fixed full-weight Pauli string $P_i \in \mathcal{S}_N$, the number of target strings $P_j$ satisfying $D_H(P_i,P_j)\leq 2$ is $1+3N+9\binom{N}{2}$. The total number of such $P_j$ over all $3^N$ choices of $P_i$ is the following:}
		\begin{equation}
			\begin{split}
				|A| &= |\mathcal{S}_N| \times \left[ 1 + \binom{N}{1} \times 3 + \binom{N}{2} \times 3^2 \right] \\
				&= 3^N \left( 1 + 3N + 9\binom{N}{2} \right).
			\end{split}
		\end{equation}
		By symmetry, the case where $P_j \in \mathcal{S}_N$ {yields the same size} $|B| = |A|$. To avoid double counting, we subtract the intersection $|A \cap B|$, where both $P_i, P_j \in \mathcal{S}_N$. In this intersection, any altered position must strictly change to another non-identity operator (only 2 choices per position, as $I$ is excluded):
		\begin{equation}
			\begin{split}
				|A \cap B| &= 3^N \left[ 1 + \binom{N}{1} \times 2 + \binom{N}{2} \times 2^2 \right] \\
				&= 3^N \left( 1 + 2N + 4\binom{N}{2} \right).
			\end{split}
		\end{equation}
		By applying $K_{\text{res}}^{(N)} = 2|A| - |A \cap B|$, {the expression for the high-order optimization complexity} is obtained:
		\begin{equation}
			K_{\text{res}}^{(N)} = 3^N \left[ 1 + 4N + 14\binom{N}{2} \right],
		\end{equation}
		\textcolor{black}{thus, $K_{\mathrm{res}}^{(N)}$ characterizes the size of the mask-selected PTM space for an $N$-qubit system. It should be noted that a general completely positive and trace-preserving (CPTP) channel acting on a d-dimensional Hilbert space contains $d^4-d^2$ independent real degrees of freedom~\cite{Choi1975,Lindblad1976}. Therefore, for the two-qubit case, the foundational model contains $d^4-d^2=256-16=240$ independent degrees of freedom. Hence, 240 serves as the starting point for the progressive optimization in the HPO framework.}
			
		\textcolor{black}{Since directly optimizing this space becomes impractical as $N$ increases, HPO freezes all parameters learned in previous stages and only optimizes the newly introduced parameters associated with the added qubit $q_N$. For the endpoint-grown topology, the corresponding trainable parameter count is denoted by 
		\begin{equation}
		\Delta K_N=\sum_{w=1}^{N}3^w,\qquad N>2.
		\end{equation}
		}
		
		\textcolor{black}{The resulting PTM-element counts are summarized in Table~\ref{tab1}, where  $C_{\mathrm{mask}}=1-K_{\mathrm{res}}^{(N)}/\mathcal{D}_{\mathrm{full}}$ denotes the corresponding compression ratio.}

\begin{table}[htbp]
	\caption{\textcolor{black}{Comparison of the required number of PTM-entries under the CPTP constraint during 2-to-10-qubit progressive training of a 10-qubit system.}}
	\label{tab1}
	\centering
	\normalsize
	\renewcommand{\arraystretch}{1.05}
	{
	\begin{tabular*}{\columnwidth}{@{\extracolsep{\fill}}ccccc@{}}
		\hline\hline
		$N$
		& $\mathcal{D}_{\mathrm{full}}$
		& $K_{\mathrm{res}}^{(N)}$
		& $C_{\mathrm{mask}}$
		& \textcolor{black}{$\Delta K_N$} \\
		\hline
		2  & 240               & 240        & $-$      & 240    \\
		3  & 4,032             & 1,485      & 63.17\%    & 39     \\
		4  & 65,280            & 8,181      & 87.47\%    & 120    \\
		5  & 1,047,552         & 39,123     & 96.27\%    & 363    \\
		6  & 16,773,120        & 171,315    & 98.98\%    & 1,092  \\
		7  & 268,419,072       & 706,401    & 99.74\%    & 3,279  \\
		8  & 4,294,901,760     & 2,788,425  & 99.935\%   & 9,840  \\
		9  & 68,719,214,592    & 10,648,503 & 99.9845\%  & 29,523 \\
		10 & 1,099,510,579,200 & 39,621,879 & 99.9964\%  & 88,572 \\
		\hline
	\end{tabular*}
	}
\end{table}
		
		\section{Numerical Simulation and Results}
		Numerical simulations to evaluate the performance and scalability of the Hierarchical Progressive Optimization (HPO) framework are performed here. 
		
		\subsection{Simulation Setup}
		To ensure stable and extremely precise optimization, an Adam optimizer is employed and initialized with a learning rate of $\eta = 0.002$, coupled with a Cosine Annealing learning rate scheduler ($\eta_{\text{min}} = 10^{-5}$). 
		To quantify the accuracy of the noise parameter extraction during the QPT process, we use the Mean Squared Error (MSE) between the predicted expectation values and the ground-truth:
		\begin{equation}
			\text{MSE} = \frac{1}{|\mathcal{S}|} \sum_{k=1}^{|\mathcal{S}|} \left( \langle O_k \rangle_{\text{pred}} - \langle O_k \rangle_{\text{true}} \right)^2,
		\end{equation}
		where $\mathcal{S}$ represents the selected combinatorial subset of high-weight Pauli observables. Furthermore, to evaluate the algorithmic impact of the noise model, the state fidelity $\mathcal{F}$ between the target density matrix $\rho_{\text{ideal}}$ and the noise-injected density matrix $\sigma_{\text{noisy}}$ is computed {\cite{Jozsa1994}}:
		\begin{equation}
			\mathcal{F}(\rho_{\text{ideal}}, \sigma_{\text{noisy}}) = \left( \text{Tr} \sqrt{\sqrt{\rho_{\text{ideal}}} \sigma_{\text{noisy}} \sqrt{\rho_{\text{ideal}}}} \right)^2.
		\end{equation}
		
		{To make HPO training computationally tractable, we further restrict the stage-wise trainable parameter space based on \textcolor{black}{$\Delta K_N$} in Table~\ref{tab1}. For $N\leq5$, the trainable count equals \textcolor{black}{$\Delta K_N$}. For $N>5$, a Pauli-weight cutoff of $w\leq3$ limits each stage to $3+9+27=39$ newly introduced parameters, while all parameters inherited from stage $N-1$ remain fixed.}
		
		{The simulations are executed on a GPU-accelerated quantum machine learning framework leveraging PennyLane \cite{Bergholm2018}, which enables high-efficiency backpropagation through complex quantum circuits.}

		\subsection{Progressive Optimization of Pauli-Weight Noise Components}
		{The learning dynamics of the HPO framework under varying qubit scales $N \in \{2, 3, 4, 5\}$ is validated firstly. The active parameter space is strictly bounded by the combinatorial progressive mask, preventing gradient dilution.}
		
		\begin{figure}[H]
			\centering
			\includegraphics[width=\columnwidth]{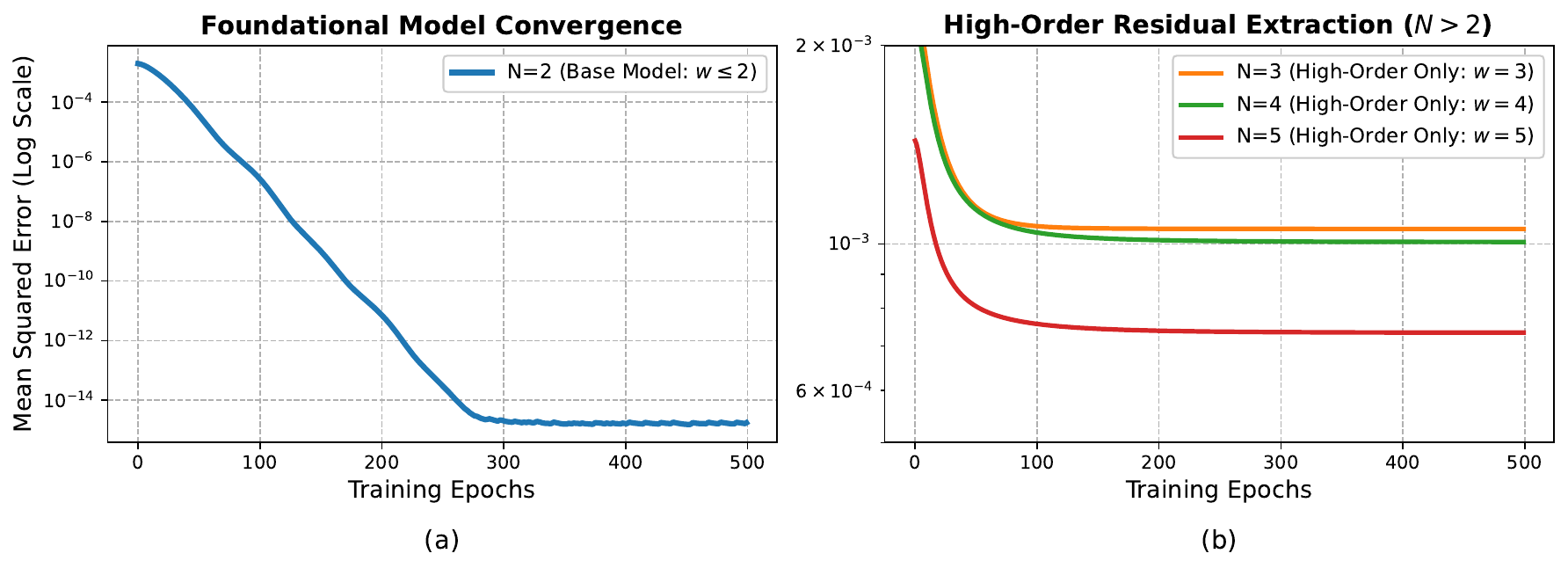}
			\caption{Convergence behavior of the HPO framework. (a) {The convergence of the foundational baseline} of the low-order terms; (b) The convergence of higher-order residual terms.}
			\label{fig:convergence}
		\end{figure}
		
		Figure \ref{fig:convergence} illustrates convergence behaviors induced by the hierarchical masking strategy of the HPO framework in different Pauli terms. As shown in Fig. \ref{fig:convergence}(a), the HPO framework, which optimizes all Pauli terms with weight $w \le 2$, exhibits a clear convergence trend, and eventually reaches an MSE of approximately $10^{-14}$. This result indicates that the dominant low-order local noise components are sufficiently represented by the $w \le 2$ Pauli interaction terms. In contrast, Fig. \ref{fig:convergence}(b) reveals the learning behavior of the higher-order residual components {with $N>2$ under the HPO framework}. After the previously optimized low-order baseline is fixed, the projection mask restricts subsequent optimization to the remaining higher-weight Pauli sectors. 
		As a result, the residual learning trend remains smooth and stable, suggesting that the HPO framework can separate the well-captured local baseline from the weaker multi-body corrections. This behavior supports the hierarchical decomposition: the foundational model first absorbs the dominant local Pauli error structure, while the residual stages focus on higher-order correlations without destabilizing the already optimized baseline.
		
		{		
		It should be emphasized that the larger MSEs in Fig.~\ref{fig:convergence}(b) do not imply reduced fidelity. In HPO, the low-order components are fixed before optimizing the weak high-order residuals. Although their small amplitudes make them more difficult to estimate, their limited contribution to the overall noise channel leaves the algorithm-level fidelity essentially unchanged.
		}

		\subsection{Quantum Error Mitigation on 10-Qubit HHL Algorithm}
		
\begin{figure}[H] %
		\centering
		\includegraphics[width=\columnwidth]{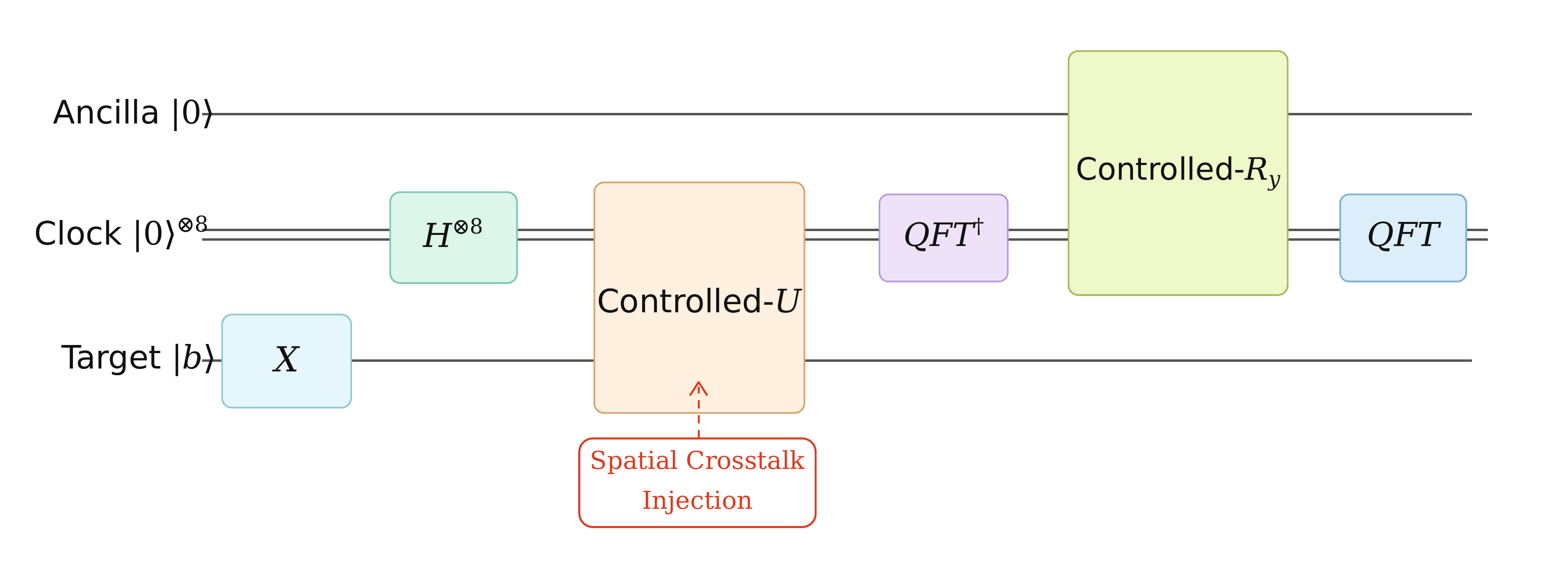}
		\caption{Quantum circuit architecture of the 10-qubit Harrow-Hassidim-Lloyd (HHL) algorithm used for spatial crosstalk mitigation. The circuit comprises 1 ancilla qubit, 8 clock qubits (denoted by the double-line bus) for high-precision quantum phase estimation, and 1 target qubit. The red dashed arrow indicates the injection points of unmitigated dynamic multi-body crosstalk during dense controlled-phase operations.}
		\label{fig_circuit}
	\end{figure}

			To demonstrate the effectiveness of the HPO framework, the proposed method is evaluated through {a circuit-level numerical simulation of} the 10-qubit Harrow--Hassidim--Lloyd (HHL) algorithm~\cite{Harrow2009}, which requires accurate quantum phase estimation (QPE) and is highly sensitive to unmitigated phase accumulation and high-order crosstalk.
		As depicted in the circuit architecture of Fig.~\ref{fig_circuit}, the simulation uses 1 ancilla qubit, 8 clock qubits for precision QPE, and 1 target qubit. {Four benchmarking settings are evaluated in terms of the target-state fidelity $\mathcal{F}$:} an ideal noiseless state, the unmitigated noisy simulation, QEM via the traditional global depolarizing model, and QEM via the HPO framework.
		
		\begin{figure}[H]
			\centering
			\includegraphics[width=0.88\columnwidth]{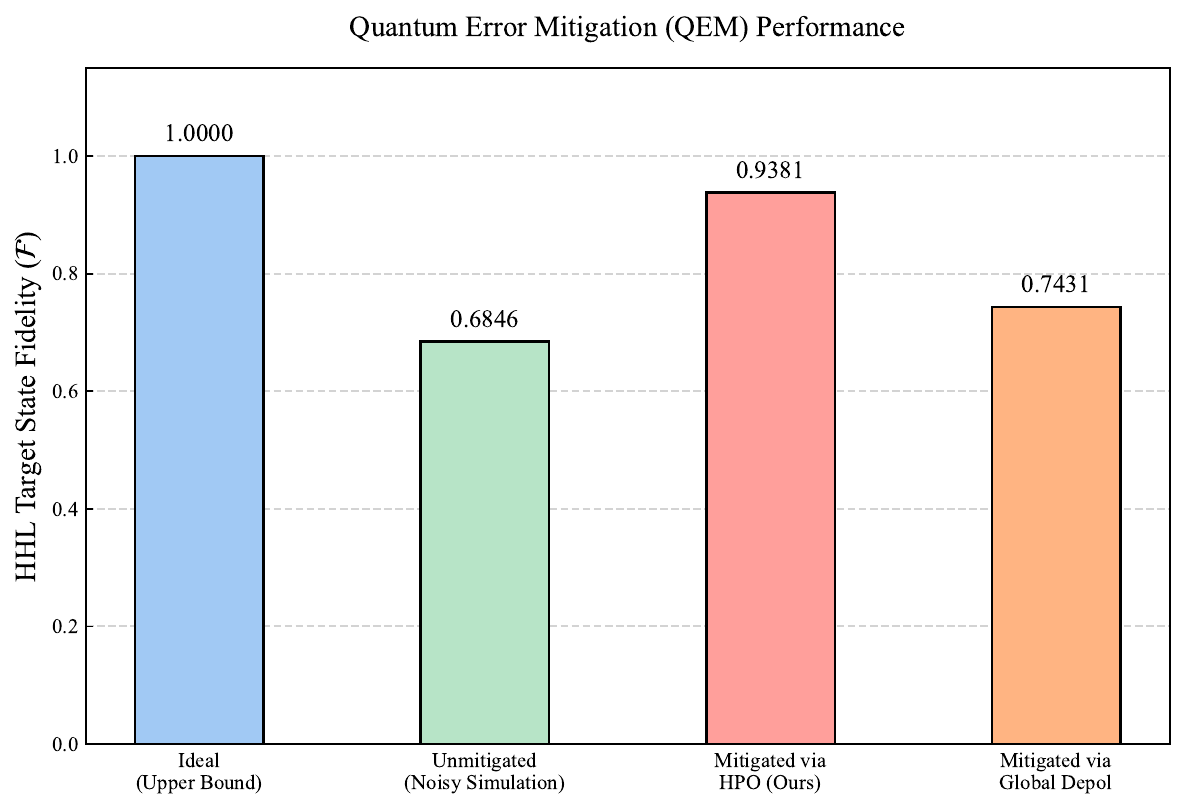}
			\caption{Algorithmic state fidelity evaluation of the 10-qubit HHL algorithm under various Quantum Error Mitigation (QEM) strategies.} 
			\label{fig:hhl_fidelity}
		\end{figure}
		
		As quantitatively illustrated in Fig. \ref{fig:hhl_fidelity}, the HPO-guided mitigation scheme targets and cancels these fragile high-weight errors. 
		Consequently, HPO-guided mitigation recovers the target-state fidelity to $0.9381$, corresponding to {an absolute fidelity gain of $19.5$ percentage points relative to the global depolarizing baseline}.
		This result suggests that sparse high-order Pauli residuals can provide actionable information for algorithm-level error mitigation, especially when the dominant noise contains structured crosstalk components. 
		In contrast, the {unmitigated noisy simulation} suffers from severe decoherence, yielding an unacceptable fidelity of $0.6846$. 
		Applying the global depolarizing mitigation baseline improves the target-state fidelity from 0.6846 to 0.7431, where the improvement remains limited, for the reason that the traditional model relies on the oversimplified assumption of independent single-qubit error channels, completely blinding itself to the non-local multi-body correlations (e.g., $XX$ or $ZZ$ crosstalk) that aggressively corrupt the deep QPE circuit.

\subsection{Scalability and Parameter Overhead}
			{
			To evaluate the scalability of the HPO framework, \textcolor{black}{its PTM-element and physical-coefficient counts are compared with the required PTM-element count} associated with Quantum Process Tomography (QPT).
		}

\begin{figure}[H]
				\centering
				\includegraphics[width=0.96\columnwidth]{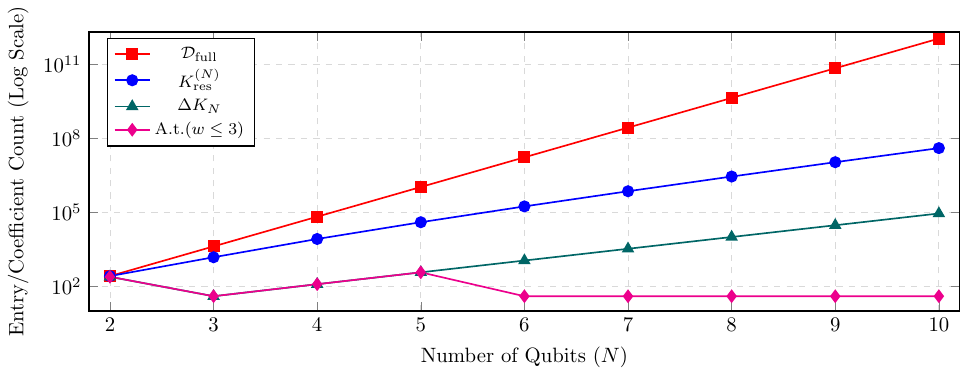}
			\caption{{Log-scale comparison of the \textcolor{black}{required PTM-element count, theoretical mask-selected PTM-element count}, untruncated incremental physical-coefficient count, and actually trainable parameters count with the Pauli-weight cutoff $w\le3$ imposed for $N>5$.}}
			\label{fig_scaling}
		\end{figure}

		{
			Figure~\ref{fig_scaling} distinguishes four counting quantities of $log$ up to $N=10$. As shown in it, the \textcolor{black}{required PTM-element count} $\mathcal{D}_{\mathrm{full}}$ exhibits the fastest exponential growth as the number of qubits $N$ increases. The theoretical \textcolor{black}{mask-selected PTM-element count} $K_{\mathrm{res}}^{(N)}$ also grows rapidly, but remains smaller than $\mathcal{D}_{\mathrm{full}}$. The \textcolor{black}{untruncated incremental physical-coefficient count $\Delta K_N$} increases more moderately, although its training dimension still grows continuously with $N$. }
			
			{
			In contrast, the stage-wise trainable coefficient count under the Pauli-weight cutoff $w\leq3$ remains constant for $N\geq6$, demonstrating that the truncation prevents the optimization dimension from increasing with the system size.
			This truncation strategy substantially reduces the stage-wise training dimension. As shown in Table~\ref{tab1}, even under the \textcolor{black}{$\Delta K_N$} scheme, $88{,}572-29{,}523=59{,}049$ newly introduced parameters must be optimized at $N=10$. This scale still imposes a considerable optimization burden on classical computing platforms. Moreover, for $N>10$, the fidelity improvement obtained by including additional high-order Pauli coefficients decreases rapidly. Therefore, optimizing all untruncated incremental parameters is both computationally expensive and inefficient, motivating the use of a Pauli-weight cutoff to limit the number of trainable parameters at each stage.
		}

		\section{Conclusion}
		In summary, we have developed and numerically verified the Hierarchical Progressive Optimization (HPO) framework, addressing {the exponential $\mathcal{O}(16^N)$ scaling of the \textcolor{black}{required PTM-element count}} in quantum noise characterization. Our method introduces a deterministic combinatorial projection mask that strictly isolates $N$-th order spatial crosstalk residuals while freezing foundational low-weight topologies. This mathematical decoupling fundamentally prevents gradient dilution and avoids barren plateaus, significantly reducing the parameter space overhead. 
		
		The practical superiority of the HPO framework is demonstrated through an algorithmic-level Quantum Error Mitigation (QEM) simulation on a 10-qubit HHL algorithm. By precisely targeting fragile, high-weight Pauli correlations (such as $XX$ and $ZZ$ crosstalk) that are routinely overlooked by traditional simplified depolarizing models, the mitigation strategy achieves a definitive fidelity recovery to $0.9381$, corresponding to {an absolute fidelity gain of approximately $19.5$ percentage points relative to the global depolarizing baseline}. 
		
		Looking forward, the HPO framework's ability to efficiently map non-local multi-body correlations opens new avenues for characterizing dynamic noise in deeper variational quantum algorithms and large-scale superconducting processors. Integrating this progressive masking technique with hardware-efficient topological error-correcting codes remains a promising direction for future investigation.
		
		\addcontentsline{toc}{chapter}{Appendix A: Appendix section heading}

		\section*{Data availability statement}
		The data that support the findings of this study are openly available. {The source code for this work is available at:} https://github.com/gexiangyu0608/HPO-QNC
		
		\addcontentsline{toc}{chapter}{Acknowledgment}
		\section*{Acknowledgment}
		This work was supported by the National Natural Science Foundation of China (No. 62375140) {and the National Innovation and Entrepreneurship Training Program for College Students (No. 202610293069Z)}.

		\addcontentsline{toc}{chapter}{References}
		\bibliographystyle{iopart-num}
		{\color{black}\bibliography{referencecpb}}
		
		
	\end{CJK*}  
\end{document}